\documentclass[12pt,a4paper,english,superscriptaddress,aps,preprint]{revtex4}
\usepackage{amsmath}
\usepackage{amssymb}
\usepackage{graphicx}
\usepackage{hhline}
\usepackage{mwe}
\usepackage{textcomp}
\usepackage{slashed}
\usepackage{geometry}
\usepackage{subfig}

\makeatletter
\usepackage{babel}
\newcommand{\bea}{\begin{eqnarray}}
\newcommand{\eea}{\end{eqnarray}}

\newcommand{\pa}{\partial}

\newcommand{\be}{\begin{equation}}
\newcommand{\ee}{\end{equation}}
\numberwithin{equation}{section}

\begin{document}
\immediate\write16{<<WARNING: LINEDRAW macros work with emTeX-dvivers
                    and other drivers supporting emTeX \special's
                    (dviscr, dvihplj, dvidot, dvips, dviwin, etc.) >>}

\title{\boldmath Non-metric construction of spacetime defects}
\preprint{KA-TP-26-2019}
\date{December 21, 2019}
\author{Jose Queiruga}
\email{jose.queiruga@kit.edu}
\affiliation{Institute for Theoretical Physics, Karlsruhe Institute
of Technology (KIT), 76128 Karlsruhe, Germany}
\affiliation{Institute for Nuclear Physics, Karlsruhe Institute of Technology (KIT), Hermann-von-Helmholtz-Platz 1, 76344 Eggenstein-Leopoldshafen, Germany}
\email{jose.queiruga@kit.edu}

\begin{abstract}
We describe a spacetime endowed with a non-metricity tensor which effectively serves as a model of a spacetime foam. We explore the consequences of the non-metricity in several $f(R)$ theories. 
 
\end{abstract}

\maketitle


\section{Introduction}
\label{intro}

It has been argued  by Wheeler  \cite{Wheeler1,Wheeler2}  that the spacetime may have a nontrivial structure over small scales. This nontrivial structure is characterized by fluctuations of the metric and the topology. The topological changes are usually referred to as spacetime defects (or spacetime foam). These ``imperfections" can be seen as remnants of a possible quantum phase of the spacetime.

Since not too much is known about the aforementioned small structure of spacetime, there is a large degree of arbitrariness in its description. Different approaches have been proposed in the study of the spacetime defects, but basically in the single defect case and for particular topologies \cite{Hawking1, Hawking2, Klinkhamer3, Klinkhamer2, Queiruga1, Queiruga2}, (for other approaches see \cite{Klinkhamer1, Queiruga4, Queiruga3}). It is not difficult to convince oneself that a spacetime with changing topology over small scales is hard to describe or even intractable. It seems, therefore, natural to take an effective perspective, where the spacetime foam is described by an intrinsic geometrical property of the spacetime manifold.  

Our point of view is inspired by the theory of defects in solids, where a solid (e.g., a crystal) can be described by a three-dimensional Riemannian manifold and the presence of point-like defects can be modeled through the non-metricity tensor, i.e., the failure of the metric to be covariantly conserved \cite{Kupferman, Gunther}. This is rather natural, since in the presence of defects one should expect a non-conservation of the volume element, which is a consequence of the non-conservation of the metric. In other words, under this interpretation non-metric theories of gravity can be used to describe the spacetime foam.

 In previous works the non-metricity was used to describe such a distribution of defects, but with matter acting as its source \cite{Hossenfelder, Latorre}. From our point of view, the spacetime foam (and hence the non-metricity) should be an intrinsic property of the spacetime and therefore one must expect its existence even in the vacuum. We will show that this problem can be overcome by giving dynamics to the non-metricity field which describe the defects. In addition, we will show that, higher-curvature corrections in non-metric gravity are consistent with Starobinsky-like models and can drive an inflationary phase and late-time accelerated expansion depending on the sign of the kinetic term of the defects.
 
 This paper is organized as follows. In Sec. \ref{nonmetricity}, the concept of non-metricity is introduced and its connection with spacetime defects is suggested. In Sec. \ref{action}, we review some results about non-metric $f(R)$ theories. In Sec. \ref{defGR} the dynamical term associated to the non-metric tensor is introduced. In Sec. \ref{sec:FRW}, explicit solutions are analyzed in a RW universe. Finally, Sec. \ref{sec:conc} contains  conclusions and further discussion. We also add one appendix with technical details about the stability of the solutions.


\section{Non-metricity and defects}
\label{nonmetricity}

The presence of point-like defects in a solid can be identified with the non-conservation of the volume element and the distribution can be described in a natural geometrical way with the introduction of the non-metricity tensor \cite{Kupferman, Gunther,Hossenfelder}. This can be seen as follows. The non-metricity tensor, $Q_{\mu\nu\rho}$, measures the failure of the metric  to be covariantly conserved,
\be
\widetilde{\nabla}_\mu g_{\nu\rho}=-Q_{\mu\nu\rho}.\label{nonm-q}
\ee
If we assume that the torsion tensor vanishes, the relation (\ref{nonm-q}) results in a modification of the connection $\widetilde{\Gamma}^\kappa_{\,\,\nu\mu}$ that can be written as the Levi-Civita connection $\Gamma^\kappa_{\,\,\nu\mu}$ plus an extra tensor:
\be
\widetilde{\Gamma}^\kappa_{\,\,\nu\mu}=\Gamma^\kappa_{\,\,\nu\mu}+\Omega^\kappa_{\,\,\nu\mu}.\label{nonm-con}
\ee

From the torsion-free condition it follows that $\Omega^\kappa_{\,\,[\nu\mu]}=0$. The (general) covariant derivative $\widetilde{\nabla}$ is now defined in terms of the new connection. The non-conservation of the volume element follows immediately since
\be
\frac{1}{g}\widetilde{{\nabla}}_\lambda \,g=-2 \Omega^\rho_{\,\, \lambda\rho}, 
\ee
 where we have defined $g=\det g_{\mu\nu}$. By keeping the analogy with the geometric description of defects in solids, the tensor $\Omega$, or rather a contraction of it, will describe a random distribution of spacetime defects through a smooth spacetime manifold.

\subsection{The single defect case}

Although we will use this non-metric approach to describe an effective distribution of defects through spacetime with its own dynamics, this could still be used to model a single static defect. For this purpose, we assume that  the $\Omega$ tensor (or the Weyl vector defined in Sec. 4.1), has support over a compact region of space, say a sphere $S_R$ of radius $R$,
\be
\Omega^\mu_{\nu\rho}(x)=
\begin{cases}
0,\,x\notin S_R ,\\
\neq 0,\,x\in S_R.
\end{cases}
\ee

The radius $R$ can be identified with the size of the defect. Of course, the support $S_R$, can be thought as a disjoint union of compact regions $S_i$, and in this case  $\Omega$ describes a distribution of defects of (possibly) different size.

 Now, we can wonder about the effect of this non-vanishing non-metricity on the trajectory of a test particle. First, it is clear that outside the region $S_R$, the connection is Levi-Civita and the geodesic equation is the usual one. However, in the region $S$, the geodesics satisfy  
\be\label{eq: geodesic equation}
\frac{d^2 x^{\mu}}{d\tau ^2} + \left(\Gamma^{\mu} _{\hphantom {1} \rho \sigma} +\Omega^{\mu} _{\hphantom {1} \rho \sigma} \right)\frac{dx^{\rho}}{d \tau}\frac{dx ^{\sigma}}{d \tau} =0 \,,\,\, x\in S_R.
\ee

The modification of the geodesic when crossing the defect (the region $S_R$) depends entirely on the specific choice of $\Omega$. In an exact description of a spacetime defect (represented by nontrivial topology in the region $S_R$), this modification is entirely determined by the boundary conditions of the surface of $S_R$ (see for example \cite{Klink-Wang}). On the one hand, it is not very difficult to convince oneself that, within this point of view, a judicious choice of $\Omega$ would allow to mimic the effects of the nontrivial topology. On the other hand, the advantage of this approach is that it could be easily extended to effectively describe a distribution of spacetime defects by choosing  $\Omega$ appropriately. Note also that the same idea can be used to describe a non-static defect by allowing $\Omega$ to depend on time. We will, however not insist on this line of reasoning. As we will see, in the Palatini formalism, $\Omega$ (and therefore the defects) will be considered a dynamical entity obeying corresponding field equations.


\section{Non-metric $f(R)$ theories}
\label{action}

If non-metricity is the main ingredient in the effective description of the presence of defects in the spacetime manifold, one has to look for actions compatible with non-vanishing  $Q_{\mu\nu\rho}$. In the Palatini formalism the metric and the connection are considered as two independent geometrical quantities (see for example \cite{Sotiriou1, Sotiriou2,Sotiriou3} and references therein). This approach fits quite well to our purposes, as $\Omega$ itself will be promoted to a genuine dynamical entity. As a consequence, in order to obtain the field equations one has to perform variations with respect to both fields. The variation with respect to $g_{\mu\nu}$ gives the Einstein field equations, while the variation with respect to $\widetilde{\Gamma}^\kappa_{\,\,\nu\mu}$ (Palatini variation) gives (possibly) some constraints on the connection coefficients. It is well-known that, if we start with the Einstein-Hilbert (EH) action and with a general connection (non-metric) the Palatini variation constrains the connection to be the Levi-Civita one. From this point of view it seems, therefore, that the presence of defects described by a non-metricity tensor requires a more general action. A natural generalization is given by $f(R)$ gravity, where the linear EH action is replaced by a general function of the Ricci scalar.


\subsection{Non-metric $f(R)$ gravity without matter}
\label{fRvac}

Let us assume that a general non-metric $f(R)$ theory can describe defects distributed on the spacetime manifold. We have the following action in vacuum
\be
S=\frac{1}{2\kappa}\int d^4x\sqrt{-g}f(R),
\ee
with $\Gamma^\mu_{\,\,[\nu\rho]}=0$ and $\widetilde{\nabla}_\mu g_{\nu\rho}=-Q_{\mu\nu\rho}\neq 0$. The metric and Palatini variations of the action lead to the following equations
\bea
&&f'(R)R_{\mu\nu}-\frac{1}{2}f(R)g_{\mu\nu}=0,\label{fRvac-eom1}\\
&&\widetilde{\nabla}_\lambda\left(f'(R)\sqrt{-g}g^{\mu\nu}\right)-\widetilde{\nabla}_\sigma\left(f'(R)\sqrt{-g}g^{\sigma(\mu}\right)\delta^{\nu)}_{\,\, \lambda}=0\label{fRvac-eom2}.
\eea
After taking the trace in (\ref{fRvac-eom1}) and (\ref{fRvac-eom2}) one obtains
\bea
&&f'(R)R-2 f(R)=0\label{fRvac-trac1}\\
&& \widetilde{\nabla}_\lambda\left(f'(R)\sqrt{-g}g^{\mu\nu}\right)=0.\label{fRvac-trac2}
\eea
Assuming that $f(R)\neq \alpha R^2$, the solutions of (\ref{fRvac-trac1}) correspond to constant Ricci scalar, $R={c_i}$ (for details see \cite{Sotiriou1}). This implies that $f(R)$ is also a constant and as a consequence
\be
\widetilde{\nabla}_\lambda\left(\sqrt{-g}g^{\mu\nu}\right)=0\Rightarrow Q_{\mu\nu\rho}=0.\label{fRvac-metric}
\ee 
Therefore, a general non-metric $f(R)$ theory corresponds to the Einstein equations with a cosmological constant
\be
R_{\mu\nu}-\frac{1}{2}c_i g_{\mu\nu}=0.
\ee
From (\ref{fRvac-metric}) it follows that the theory is metric, and, if the non-metricity tensor describes a distribution of defects, $f(R)$ theories in vacuum cannot contain them.
As already mentioned, it is reasonable to assume that the defects are an intrinsic property of the spacetime, and hence they should exist regardless the presence  of matter.  From all these considerations we conclude that $f(R)$ gravity in vacuum does not have enough structure to describe defects and therefore, we will explore general actions providing dynamics to the defects.


\section{New action for defects and gravity}
\label{defGR}

The considerations above suggest that, in order to describe spacetime defects, one has to consider more general actions than $f(R)$ gravity.  We will be interested in actions of the form
 \be
 S=\frac{1}{2\kappa}\int d^4 x\sqrt{-g}\left(f(\tilde{R})+P[\pa_\sigma \Omega^\mu_{\,\, \nu\rho}]\right), \label{defGR-act}
 \ee
where $P$ is at most quadratic in $\pa_\sigma \Omega^\mu_{\,\, \nu\rho}$ and responsible for the dynamics of $\Omega^\mu_{\,\, \nu\rho}$ and $\widetilde{R}$ is defined in (\ref{eq:ricci tensor with traceless omega}). But before doing that, and in order to see the effect of the non-metricity, let us consider a simple situation contained in (\ref{defGR-act})
\be
S=\frac{1}{2\kappa}\int  d^4x \sqrt{-g}\tilde{R}.\label{act-nme}
\ee

The Ricci tensor including non-metricity can be expanded as follows
\begin{equation}\label{eq:ricci tensor in general}
\begin{aligned}
R_{\mu \sigma}(\widetilde{\Gamma})  & =  \partial_{\nu} \widetilde{\Gamma } ^{\nu}_{\hphantom {1} \mu \sigma }-\partial_{\mu} \widetilde{\Gamma } ^{\nu}_{\hphantom {1} \nu \sigma } +\widetilde{\Gamma } ^{\lambda}_{\hphantom {1} \mu \sigma } \widetilde{\Gamma } ^{\nu}_{\hphantom {1} \lambda \nu } -\widetilde{\Gamma } ^{\lambda}_{\hphantom {1} \nu \sigma } \widetilde{\Gamma } ^{\nu}_{\hphantom {1} \lambda \mu } \\
                               & =R_{\mu \sigma}(\Gamma)+R_{\mu \sigma} (\Omega) + \Omega ^{\nu}_{\hphantom {1} \nu \lambda } \Gamma ^{\lambda}_{\hphantom {1}  \mu \sigma } +\Omega ^{\lambda}_{\hphantom {1} \mu \sigma } \Gamma ^{\nu}_{\hphantom {1}  \lambda \nu }-\Omega ^{\lambda}_{\hphantom {1} \nu \sigma } \Gamma ^{\nu}_{\hphantom {1}  \mu \lambda }-\Omega ^{\nu}_{\hphantom {1} \mu \lambda } \Gamma ^{\lambda}_{\hphantom {1}  \nu \sigma } \,,
\end{aligned}
\end{equation}
where 
\begin{equation}
R_{\mu \sigma} (\Omega) \equiv  \partial_{\nu} \Omega ^{\nu}_{\hphantom {1} \mu \sigma }-\partial_{\mu} \Omega ^{\nu}_{\hphantom {1} \nu \sigma } +\Omega ^{\lambda}_{\hphantom {1} \mu \sigma } \Omega ^{\nu}_{\hphantom {1} \lambda \nu } -\Omega ^{\lambda}_{\hphantom {1} \nu \sigma } \Omega ^{\nu}_{\hphantom {1} \lambda \mu } \, .
\end{equation}
If we perform the variation with respect to $\Omega^\mu_{\,\, \nu\rho}$ in (\ref{act-nme}) and trace in the $\rho,\sigma$ indices and contract in the $\rho,\lambda$ indices,  the following constraints in the non-metric part of the connection are imposed
\be
\Omega^\mu_{\,\,\mu\nu}=0,\quad g^{\mu\nu}\Omega^\rho_{\,\,\mu\nu}=0.\label{tracond}
\ee

Taking into account the conditions (\ref{tracond}) the Ricci scalar can be simplified as 
\begin{equation}\label{eq:ricci tensor with traceless omega}
\widetilde{R}\equiv R(\widetilde{\Gamma} ,g_{\mu \nu})=g^{\mu \sigma}R_{\mu \sigma}(\widetilde{\Gamma}) = R(\Gamma , g_{\mu \nu} ) - \Omega _{\mu \lambda \nu} \Omega ^{\lambda \mu \nu} 
\end{equation} 

Upon inserting (\ref{eq:ricci tensor with traceless omega}) in (\ref{act-nme}) we obtain
\begin{equation}\label{eq:EH+lambda}
S=\frac{1}{2\kappa}\int d^4x\sqrt{-g}\left\lbrace R(\Gamma , g_{\mu \nu} )-\Omega _{\mu \lambda \nu} \Omega ^{\lambda \mu \nu} \right\rbrace \,.
\end{equation}

Therefore, once we consider a non-metric EH action, a mass term (of Planck order) for the non-metricity tensor is naturally generated. Moreover, $\Omega _{\mu \lambda \nu} $ is trivially eliminated from (\ref{eq:EH+lambda}) generating the standard EH action (see \cite{Burton} for other non-metric extensions of the EH action).


\subsection{Weyl vector and the addition of dynamics}

So far we have considered a general non-metricity tensor with (potentially) 40 degrees of freedom. From now on we will assume that all relevant d.o.f. of $\Omega^\mu_{\nu\rho}$ are contained in a vector $W_\mu$ (generally called Weyl vector) defined as follows 
\begin{equation}\label{eq:omega of W}
\Omega ^{\nu}_{\hphantom {1} \mu \sigma} = \frac{1}{2}(W_{\sigma} \delta^{\nu} _{\mu}+W_{\mu} \delta ^{\nu} _{\sigma} - W^{\nu} g_{\mu \sigma}) ,
\end{equation}
or
\be
W_\mu=\frac{1}{2}\Omega^\nu_{\,\,\nu\mu}.
\ee
The Ricci tensor $R_{\mu \sigma} (\widetilde{\Gamma})$ can be written as 
\begin{equation}\label{eq:ricci tensor of W}
\begin{split}
R_{\mu \sigma}(\widetilde{\Gamma}) = & R_{\mu \sigma}({\Gamma}) + \frac{1}{2}\partial_{\sigma} W_{\mu}-\frac{3}{2}\partial_{\mu} W_{\sigma}-\frac{1}{2}\partial_{\nu}(W^{\nu}g_{\mu \sigma})+W_{\lambda}\Gamma ^{\lambda}_{\hphantom {1} \mu \sigma}  - \frac{1}{2} g_{\mu \sigma} W^{\lambda}\Gamma ^{\nu}_{\hphantom {1} \nu\lambda} \\ 
&
+ \frac{1}{2} W_{\mu} W_{\sigma}-\frac{1}{2}g_{\mu \sigma}W^2 + \frac{1}{2} g_{\nu \sigma}W^{\lambda}\Gamma ^{\nu}_{\hphantom {1} \mu \lambda} +\frac{1}{2} g_{\nu \mu} W^{\lambda}\Gamma ^{\nu}_{\hphantom {1} \lambda \sigma} .
\end{split}
\end{equation}

Then the EH action in terms of (\ref{eq:ricci tensor of W}) is given by 
\begin{align}\label{eq:E-H action of W}
S & =\frac{1}{2\kappa}\int d^4 x \sqrt{-g} R(\widetilde{\Gamma},g_{\mu \nu}) \nonumber\\
  & = \frac{1}{2\kappa} \int d^4 x \left\lbrace \sqrt{-g}\left[R(\Gamma ,g_{\mu \nu})-\frac{3}{2}W^2 \right]-3\partial_{\nu}(\sqrt{-g}W^{\nu})\right\rbrace,
\end{align}
where $R = g^{\mu \nu} R_{\mu \nu}$ is the Ricci scalar and the boundary term is explicitly shown. In order to obtain the second line of (\ref{eq:E-H action of W}), we have used the following identities
\begin{subequations}\label{eq:tips1}
\begin{eqnarray}
g^{\mu \sigma} \partial_{\nu} g_{\mu \sigma} &=& 2  \Gamma ^{\lambda} _{\hphantom {1} \nu \lambda}\,,  \\[2mm]
g^{\mu \sigma} \partial_{\mu} g_{\nu \sigma}&=&\Gamma ^{\lambda} _{\hphantom {1} \nu \lambda}+ g^{\mu \sigma} g_{\lambda \nu}   \Gamma ^{\lambda} _{\hphantom {1} \mu \sigma}\,,   \\[2mm]
\partial _{\nu} \sqrt{-g} &=& \sqrt{-g}\Gamma ^{\lambda} _{\hphantom {1} \nu \lambda}\,,
\end{eqnarray}
\end{subequations}

Note that the effect of the Weyl vector in the standard EH is the introduction of a quadric term in $W_\mu$. The variation of the EH action with respect to the Weyl vector gives 
\begin{equation}\label{eq:eom-W-EH}
-3\sqrt{-g}W^{\mu}=0.
\end{equation}
Therefore, as already stated, the non-metricity vanishes in vacuum. A natural candidate to make $W_\mu$ nontrivial is a Maxwell-like term 
\begin{equation}\label{eq:action-W-kinetic} 
L_1 = -\frac{1}{4}\sqrt{-g} g^{\mu \alpha}g^{\nu \beta}(\widetilde{\nabla} _{\mu}W_{\nu}-\widetilde{\nabla} _{\nu}W_{\mu})(\widetilde{\nabla} _{\alpha}W_{\beta}-\widetilde{\nabla} _{\beta}W_{\alpha}).
\end{equation}

The variation of the action corresponding to (\ref{eq:action-W-kinetic}) with respect to the Weyl vector gives 
\begin{align}\label{eq:eom-W-alone}
0 & =\widetilde{\nabla} _{\mu}[\sqrt{-g}g^{\mu \alpha}g^{\nu \beta}(\widetilde{\nabla} _{\alpha}W_{\beta}-\widetilde{\nabla} _{\beta}W_{\alpha})]\nonumber \\
  & =\sqrt{-g}g^{\mu \alpha}g^{\nu \beta}\widetilde{\nabla}_{\mu}(\widetilde{\nabla} _{\alpha}W_{\beta}-\widetilde{\nabla} _{\beta}W_{\alpha})+ (\widetilde{\nabla} _{\alpha}W_{\beta}-\widetilde{\nabla} _{\beta}W_{\alpha})\widetilde{\nabla}_{\mu}(\sqrt{-g}g^{\mu \alpha}g^{\nu \beta})\nonumber \\
  & =\sqrt{-g}g^{\mu \alpha}g^{\nu \beta}\widetilde{\nabla}_{\mu}(\widetilde{\nabla} _{\alpha}W_{\beta}-\widetilde{\nabla} _{\beta}W_{\alpha}) \,.
\end{align}
 Together with the EH action, we can obtain the field equation of the Weyl vector 
\begin{equation}\label{eq:eom-W}
g^{\mu \alpha}g^{\nu \beta}\widetilde{\nabla}_{\mu}(\widetilde{\nabla} _{\alpha}W_{\beta}-\widetilde{\nabla} _{\beta}W_{\alpha})=3W^{\nu}.
\end{equation}

The term in the r.h.s of (\ref{eq:eom-W}) comes from the EH action and corresponds to a ``mass" term for $W_\mu$. The full EH action can be cast as
\be\label{eq:action-EH+W-kinetic}
S =\frac{1}{2\kappa}\int d^4 x \sqrt{-g} \left\lbrace R(\widetilde{\Gamma},g_{\mu \nu}) -\frac{1}{4\lambda}  g^{\mu \alpha}g^{\nu \beta}(\widetilde{\nabla} _{\mu}W_{\nu}-\widetilde{\nabla} _{\nu}W_{\mu})(\widetilde{\nabla} _{\alpha}W_{\beta}-\widetilde{\nabla} _{\beta}W_{\alpha})  \right\rbrace \,,
\ee
where $\lambda$ is a coupling constant of dimension $[L]^{-2}$ . The variation of action (\ref{eq:action-EH+W-kinetic}) with respect to Weyl vector  and metric gives the following equations of motion respectively 
\be\label{eq:W-eom-EH+W-kinetic}
\frac{1}{\lambda} g^{\mu \alpha}g^{\nu \beta}\widetilde{\nabla}_{\mu}F_{\alpha \beta}=3W^{\nu}\,,
\ee
\be\label{eq:metric-eom-EH+W-kinetic}
R_{\mu \nu} (\Gamma) - \frac{1}{2} g_{\mu \nu} R(\Gamma ,g_{\mu \nu})= \frac{1}{2\lambda} g^{\rho \alpha} F_{\mu \rho}F_{\nu \alpha}-\frac{1}{8\lambda} g_{\mu \nu}g^{\rho \alpha} g^{\sigma \beta}F_{\rho \sigma}F_{\alpha \beta}+\frac{3}{2}W_{\mu}W_{\nu} -\frac{3}{4}g_{\mu \nu} W^2 \,,
\ee
where 
\be\label{eq:def-F}
F_{\alpha \beta} \equiv (\partial_{\alpha}W_{\beta}-\partial _ {\beta}W_{\alpha}).
\ee

First, for $1/\lambda\rightarrow0$, the non-metricity disappears and  (\ref{eq:W-eom-EH+W-kinetic}) and (\ref{eq:metric-eom-EH+W-kinetic}) reduce to the standard Einstein equations in vacuum. Second, it can be shown that the field equation for $W_\mu$ reduces to 
\be\label{eq:W-eom-EH+W-kinetic2}
\frac{1}{\lambda} g^{\mu \alpha}g^{\nu \beta}{\nabla}_{\mu}F_{\alpha \beta}=3W^{\nu}\,,
\ee
i.e. $\widetilde{\nabla}$ has been replaced by $\nabla$ (which preserves the metric). Since $[\nabla _\nu,\nabla _\mu]\propto R_{\mu\nu}$, (\ref{eq:W-eom-EH+W-kinetic2}) implies the following conservation law
\be
\nabla _\mu W^{\mu}=0 \,\label{conservation1}.
\ee
 Note that the l.h.s. of (\ref{eq:W-eom-EH+W-kinetic2}) depends on $W_\mu$ through the normal covariant derivative and as a consequence it can be interpreted as the standard Proca equation in curved spacetime.


\subsection{Conservation of energy-momentum tensor}

Let us consider that the Lagrangian for matter takes the general form $\sqrt{-g}L_M (g_{\mu \nu}, W^{\mu},\psi)$, where $\psi$ collectively denotes the matter fields. Then, the total action is given by
\be\label{eq:action_EH+W+matter}
S = \frac{1}{2\kappa} \int d^4 x \sqrt{-g} \left\lbrace R(\widetilde{\Gamma},g_{\mu \nu}) -\frac{1}{4\lambda}  g^{\mu \alpha}g^{\nu \beta}F_{\mu \nu} F^{\alpha \beta} \right\rbrace +\int d^4 x \sqrt{-g} L_M \,.
\ee

The variation of (\ref{eq:action_EH+W+matter}) with respect to Weyl vector and metric gives the following equations of motion respectively 
\be\label{eq:W-eom-EH+W+M}
{\nabla}_{\mu}F^{\mu \nu}=3\lambda W^{\nu}- 2\kappa\lambda\frac{\delta L_M}{\delta W_\nu} \,,
\ee
\be\label{eq:metric-eom-EH+W+M}
G_{\mu \nu}= \frac{1}{2\lambda} g^{\rho \alpha} F_{\mu \rho}F_{\nu \alpha}-\frac{1}{8\lambda} g_{\mu \nu}F_{\rho \sigma}F^{\rho \sigma}+\frac{3}{2}W_{\mu}W_{\nu} -\frac{3}{4}g_{\mu \nu} W^2 +\kappa T_{\mu \nu}\,,
\ee
where $T_{\mu \nu}$ is the energy momentum tensor defined in the usual way 
\be\label{eq:define-energy-momentum-tensor}
T_{\mu \nu} \equiv -\frac{2}{\sqrt{-g}}\frac{\delta S_M}{\delta g^{\mu \nu}} \,.
\ee
By taking $ \nabla _\nu $ on both sides of (\ref{eq:W-eom-EH+W+M}), we get 
\be\label{eq:contrain on W and matter}
0=\nabla _{\nu} \nabla_{\mu} F^{\mu \nu} = 3\lambda\nabla _{\nu}W^{\nu}- 2\kappa\lambda\nabla _{\nu}\frac{\delta L_M}{\delta W_\nu} \,.
\ee
Meanwhile, by taking $ \nabla ^{\nu} $ on both sides of (\ref{eq:metric-eom-EH+W+M}), we get 
\be\label{eq:conservation law}
\begin{aligned}
0=\nabla ^{\nu}G_{\mu \nu} &=\frac{1}{2\lambda}F_{\mu \nu}\left(\frac{3}{h}W^{\nu}- 2\kappa\lambda\frac{\delta L_M}{\delta W_\nu}  \right)
+\frac{3}{2} \nabla ^{\nu}   (W_{\mu}W_{\nu}) -\frac{3}{4}\nabla ^{\nu}(g_{\mu \nu} W^2) +\kappa\nabla ^{\nu} T_{\mu \nu}\\
      &= \frac{3}{2}W_{\mu}\nabla _{\nu}W^{\nu}-\kappa F_{\mu \nu} \frac{\delta L_M}{\delta W_\nu} +\kappa\nabla ^{\nu} T_{\mu \nu}\\
      &=\kappa\nabla ^{\nu} T_{\mu \nu}+\kappa W_{\mu}\nabla _{\nu}\frac{\delta L_M}{\delta W_\nu}-\kappa F_{\mu \nu} \frac{\delta L_M}{\delta W_\nu} ,
\end{aligned}
\ee
where we have used (\ref{eq:contrain on W and matter}) and the Bianchi identity. A sufficient condition for the vanishing of the last two terms in (\ref{eq:conservation law}) is that 
\be\label{eq:constrain on matter action}
 \frac{\delta L_M}{\delta W_\nu}=0 \,,
\ee
i.e., as long as the matter Lagrangian does not couple to the Weyl vector, the usual energy-momentum tensor is conserved. There are of course fields that have this property, being the obvious example a scalar. We will comment about the role of a scalar field in Sec. 6. In the presence of spinor fields, our formalism will introduce a coupling to the Weyl vector. In this case, it is still possible to define a conserved object which is a combination of the usual energy-momentum tensor and a piece associated to the defects, see for example \cite{Hossenfelder}.




\section{Some cosmological implications} 
\label{sec:FRW}

Having stablished the main properties of the non-metricity field within our approach, we move now to some  implications of the presence of non-metricity in various models. Once again, we will assume that all the relevant d.o.f. of the non-metric part of the connection are described by the Weyl vector. In particular, we will consider two non-metric $f(R)$-theories, namely a linear and a quadratic function of the Ricci scalar in addition to a Maxwell-like term giving dynamics to the Weyl vector. 
 \subsection{Non-metric $\widetilde{R}$ theory}
 \label{sub:sec:R1}

We start with spatially-flat Robertson-Walker (RW) metric
\be\label{eq:frw-metric}
 ds^2= -dt^2 + a^2 (t) \delta _{ij} dx^i dx^j \,.
\ee

As a warming-up example let us consider the action (\ref{eq:action-EH+W-kinetic}) then, the field equations for $W_\mu$ in RW spacetime can be written explicitly as 
\begin{subequations}\label{eq:FRW-eom-W}
\begin{eqnarray}
-3a^2\lambda W^0 &=&   \Delta W_0 - \partial _0 (\partial _1 W_1 +\partial _2 W_2+\partial _3 W_3)  \,,  \\[2mm]
3\lambda W_i&=&\frac{1}{a^2}\pa_j F_{ji}-\pa_0 F_{0i}-\frac{\dot{a}}{a}F_{0i},\,\, j\neq i
\end{eqnarray}
\end{subequations}
where $\Delta$ stands for Laplace operator. If we assume the ansatz, $W^{\mu} =W^{\mu}(t)$, then equations (\ref{eq:FRW-eom-W}) reduce to
\be\label{eq:FRW-eom-W0}
W^{0} =0 \,,
\ee
for the time component and 
\be\label{eq:FRW-eom-Wi}
\ddot{W}_i+\frac{\dot{a}}{a} \dot{W_i} +3\lambda W_i =0  \,.
\ee
for the spatial components. The Friedmann equations read
\bea
&&3\frac{\dot{a}^2}{a^2}=\frac{3}{4}\frac{1}{a^2\lambda}\dot{W}^2+\frac{9}{4}\frac{W^2}{a^2}\label{eq:Frid-1},\\
&&-\left(2\frac{\ddot{a}}{a}+\frac{\dot{a}^2}{a^2}\right)=\frac{1}{4}\frac{1}{a^2\lambda}\dot{W}^2-\frac{3}{4}\frac{W^2}{a^2} \label{eq:Frid-2}.
\eea

The r.h.s. of (\ref{eq:metric-eom-EH+W-kinetic}) can be interpreted as a contribution to the energy momentum tensor
\be\label{eq-EM-Weyl}
T_{\mu\nu}^{W}= \frac{1}{2\lambda} g^{\rho \alpha} F_{\mu \rho}F_{\nu \alpha}-\frac{1}{8\lambda} g_{\mu \nu}g^{\rho \alpha} g^{\sigma \beta}F_{\rho \sigma}F_{\alpha \beta}+\frac{3}{2}W_{\mu}W_{\nu} -\frac{3}{4}g_{\mu \nu} W^2,
\ee
which is non-diagonal. But the Einstein tensor is diagonal in the RW metric. This implies that, in order to solve consistently the Einstein equations, one should assume that 
\be
T_{\mu\neq\nu}^{W}= 0.\label{def:nondiag}
\ee
In order to realize this condition we proceed as follows. We replace the definition (\ref{eq:omega of W}) by

\begin{equation}\label{eq:omega of W1}
\Omega ^{\nu}_{\hphantom {1} \mu \sigma} = \frac{1}{2}\sum_{a=1}^N(W^{(a)}_{\sigma} \delta^{\nu} _{\mu}+W^{(a)}_{\mu} \delta ^{\nu} _{\sigma} - W^{(a)\,\nu} g_{\mu \sigma}) ,
\end{equation}
where $W^{(a)}$ are $N$ Weyl vectors with randomly oriented directions \cite{Mukhanov-Golovnev}. Each product $W_i^{(a)}W_i^{(a)}$ will be proportional to $N$, while the non-diagonal terms $W_i^{(a)}W_j^{(a)},\,\, i\neq j$ will be proportional to $\sqrt{N}$ (due to the random distribution of orientations). Therefore, for large $N$, the non-diagonal terms are suppressed, leading to the condition (\ref{def:nondiag}), which is exact is the limit. The same conclusion follows if one considers a triad of mutually perpendicular vectors \cite{armendariz}. In any case, and to simplify the notation, we simply assume that the non-diagonal terms are negligible, and each occurrence of $W^2$ is understood as proportional to $N$. The energy density and pressure for the Weyl tensor can be read from (\ref{eq-EM-Weyl}),
\bea
\rho&=&\frac{3}{4}\frac{1}{a^2\lambda}\dot{W}^2+\frac{9}{4}\frac{W^2}{a^2}\label{eq:rho-non},\\
p&=&\frac{1}{4}\frac{1}{a^2\lambda}\dot{W}^2-\frac{3}{4}\frac{W^2}{a^2} .\label{eq:p-non}
\eea 

In addition, if we assume that for early times the potential energy dominates, $\vert \lambda\vert W^2>>\dot{W}^2$, the equation of state of the ``Weyl fluid" takes the form
\be
p\approx-\frac{1}{3} \rho.
\ee

From the second Friedmann equation (\ref{eq:Frid-2}) we obtain the following behavior for the scale factor
\be
a(t)\propto t.
\ee
The first obvious consequence is that, non-metricity itself does not allow for an inflationary phase. The energy density for the defects $\rho_d$ defined in (\ref{eq:rho-non}) behaves as $\rho_d\propto1/t^2$ independent of the coupling constant $\lambda$. On the other hand, if we assume that the kinetic energy dominates, $ \dot{W}^2>>\vert\lambda\vert W^2$, the equation of state is given by
\be
p\approx\frac{1}{3}\rho,
\ee
which corresponds to the equation of state for radiation. Therefore, in this regime the model mimics a universe dominated by radiation, $a(t)\propto \sqrt{t}$. In this situation the energy density $\rho_d$ behaves as $\rho_d\propto\frac{1}{\sqrt{\lambda}t^{3/2}}$. As a consequence, in the aforementioned assumptions, the 3-volume multiplied by the defect energy density grows linearly in time as long as the potential energy dominates, and after that reaches a constant value determined by the coupling constant $\lambda$ and the initial values of $W$ and $a$. With this simple example we still cannot put any constraint on the coupling $\lambda$, since, as we have seen,  the only sensible quantity is the energy density of the defects, and we cannot put constraints on that either. In the next section we study a less simple example with reacher phenomenology.

 
 \subsection{Non-metric $\widetilde{R}+\alpha \widetilde{R}^2$ theory}
 \label{sub:sec:R2}
 
 We have determined that the Einstein-Hilbert action coupled to the Weyl vector cannot drive an inflationary phase. In this section we explore the effect of higher-curvature corrections. The expression for the non-metric Ricci scalar has the form (\ref{eq:E-H action of W})
\be\label{eq:Ricci nm}
R(\widetilde{\Gamma},g_{\mu \nu}) =
  \left[R(\Gamma ,g_{\mu \nu})-\frac{3}{2}W^2 \right]-\frac{3}{\sqrt{-g}}\partial_{\nu}(\sqrt{-g}W^{\nu}).
\ee

The last term in (\ref{eq:Ricci nm}) is in general a boundary term of the linear action and can be expanded as
\be\label{eq:boundary W}
\partial_{\nu}(\sqrt{-g}W^{\nu})=\Gamma^\lambda_{\,\, \nu\lambda}W^\nu+\pa_\nu W^\nu.
\ee 

In addition, in  the RW geometry we have $\Gamma^\lambda_{\,\, \nu\lambda}=0$. If we use the \textit{ansatz}, $W^0=0,W^i=W^i(t)$ then the last term in (\ref{eq:boundary W}) also vanishes. Under this conditions, the action containing a quadratic curvature term can be written as follows
\bea\label{eq:quadratic_nonme}
S_g&=&\frac{1}{2\kappa}\int d^4x\sqrt{-g}\left[\widetilde{R}+\alpha \widetilde{R}^2\right]=\nonumber\\
&&\frac{1}{2\kappa}\int d^4x\sqrt{-g}\left[R+\alpha R^2-\frac{3}{2}W^2-3\alpha R W^2+\frac{9\alpha}{4}W^4 \right].
\eea

The $\widetilde{R}^2$ term generates a quartic term in the Weyl vector and a quadratic term in the standard Ricci scalar, but also a non-minimal coupling between the Ricci scalar and the Weyl vector. After the addition of the Maxwell-like term to (\ref{eq:quadratic_nonme}) the full action that we are going to consider is
\be
S=S_g-\frac{1}{8\kappa\lambda}\int d^4x F_{\mu\nu}F^{\mu\nu}.
\ee
 The  modified Friedmann equations and the field equations for $W$ take the following form
\bea
&&H^2+36\alpha H^2\dot{H}-6\alpha \dot{H}^2+12\alpha H\ddot{H}=\frac{1}{3}\rho\equiv-\frac{1}{3}T_0^{\,\,0},\label{eq:eins1}\\
&&\dot{H}+36\alpha \dot{H}^2+6\alpha\left(3 H\ddot{H}+\dddot{H}\right)=-\frac{1}{2}(\rho+p)\equiv-\frac{1}{2}\left(-T_0^{\,\, 0}+T_i^{\,\,i}\right),\label{eq:eins2}\\
&&\ddot{w}_i+3 H \dot{w}_i+\lambda\left(\frac{1}{\lambda}+36\alpha\right)w_i\left(2H^2+\dot{H}\right)+3\lambda w_i-27\alpha\lambda w_i^3=0, \label{eq:W1}
\eea
where we have introduced the new vector field $w_i$  \cite{Mukhanov-Golovnev} defined by 
\be
 w_i=W_i/a.
\ee
The energy momentum tensor is given by
\bea
T_{\mu\nu}&=&\frac{3}{2}W_\mu W_\nu-\frac{3}{4}g_{\mu\nu}W^2+3\alpha R W_\mu W_\nu+3\alpha W^2 G_{\mu\nu}-\frac{9\alpha}{2}W_\mu W_\nu W^2+\frac{9\alpha}{8}g_{\mu\nu}W^4\nonumber \\
&+& \frac{1}{2\lambda} g^{\rho \alpha} F_{\mu \rho}F_{\nu \alpha}-\frac{1}{8\lambda} g_{\mu \nu}g^{\rho \alpha} g^{\sigma \beta}F_{\rho \sigma}F_{\alpha \beta}+3\alpha \left(g_{\mu\nu}\square-\nabla_\mu\nabla_\nu\right)W_\rho W^\rho.
\eea
From now on we will assume in addition that the field $w_i$ (including the metric factor) represents the defects (note that in flat space both definitions coincide) and $\kappa=1$. After using (\ref{eq:W1}) we can write the energy density and pressure in terms of $w_i$ in the following form
\bea
\rho&=&\frac{1}{4\lambda} \dot{w}^2+\frac{3}{4} wH\left(\frac{1}{\lambda}+36\alpha\right)\left(Hw+2\dot{w}\right)+\frac{9}{8}w^2\left(2-9w^2\alpha\right),\label{energy-den1}\\
p&=&\frac{1}{4}\left(\frac{1}{\lambda}-72\alpha\right)\dot{w}^2+\frac{\left(1+36\alpha\lambda\right)}{4}w\left(\left(\frac{1}{\lambda}+144\alpha\right)H^2w+72\alpha w\dot{H}+\frac{1}{\lambda}2H\dot{w}\right)\nonumber\\
&&-\frac{3\lambda}{8}w^2\left(2(\frac{1}{\lambda}-72\alpha)+9\alpha(\frac{1}{\lambda}+144\alpha)w^2\right),\label{press1}
\eea
where we have used the notation $w^2\equiv w_i w_i, \,\, w\dot w\equiv w_i\dot{w}_i$. In terms of the field $w_i$ it is clear that, when $\frac{1}{\lambda}+36\alpha=0$, (\ref{energy-den1}) and (\ref{press1}) can be rewritten as
\bea
\rho&=&\frac{3}{4\lambda}\dot{w}^2+\frac{9}{4}w^2+\frac{9}{32\lambda}w^4\label{EMT:limit1}, \\
p&=&\frac{3}{4\lambda}\dot{w}^2-\frac{9}{4}w^2-\frac{9}{32\lambda}w^4,\label{EMT:limit2}
\eea
while (\ref{eq:W1}) reduces to 
\be
\ddot{w}_i+3 H \dot{w}_i+3\lambda w_i-27\alpha\lambda w_i^3=0.\label{eq:w:limit}
\ee
In this limit, the model corresponds to a (metric) $R+\alpha R^2$ minimally coupled to a scalar field $\phi^2=w^2$ with potential $V(\phi)=\frac{9}{4}\phi^2+\frac{9}{32\lambda}\phi^4$.


\subsubsection{$\lambda>0,\,\, \alpha>0$}

\begin{figure}
  \centering
  \includegraphics[width=0.9\textwidth]{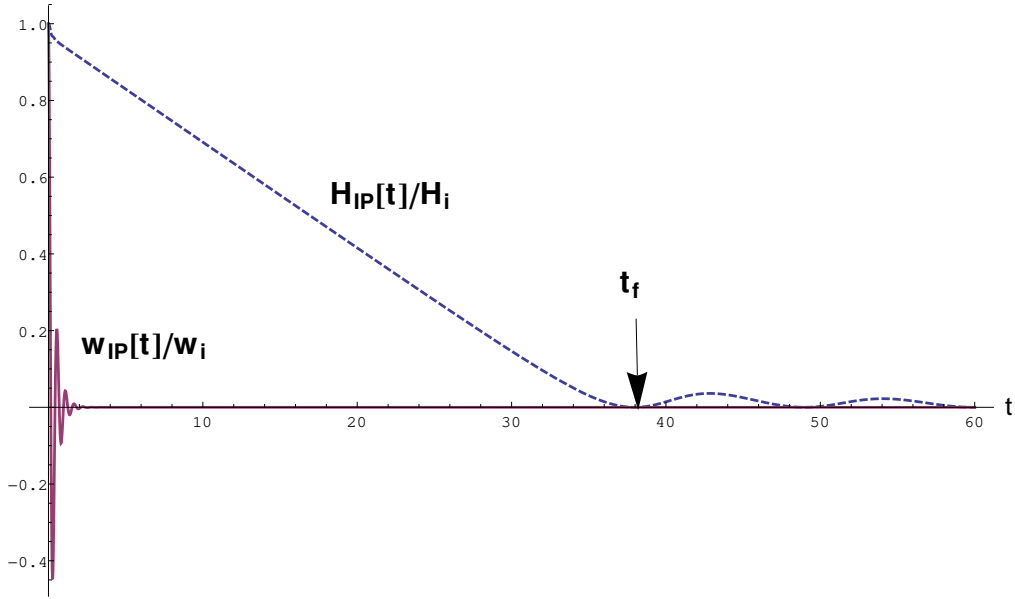}
  \caption{Normalized behavior of the Hubble parameter and the $w \,(=W/a)$ vector. $t_f$ is the time at the end of the inflationary phase and can be approximated as $t_f\approx 36H_i \alpha$.}
  \label{figpI}
\end{figure}

The situation with $\lambda>0$ and $\alpha>0$ is qualitatively similar to a metric model $R+\alpha R^2$. The field $w$ drops rapidly to zero during the inflationary phase (IP) and begins to oscillate. The universe enters in a matter dominated Friedmann phase (MFP) and the expansion decelerates forever. The Hubble parameter decays linearly in the IP and it is approximately described by the following expression
\be
H_{IP}(t)\approx H_i -\frac{1}{36 \alpha}t,\label{eq:H-IP}
\ee
where $H_i$ in the initial value if the Hubble parameter. After the IP, H(t) starts to oscillate leading to a MFP. The qualitative behavior of $H(t)$ and $w(t)$ is shown in Fig. \ref{figpI}. The number of e-folds $\mathcal{N}$ during this phase can be given by $\mathcal{N}\approx 18 H_i^2 \alpha$, as it can be seen immediately form Fig. \ref{figpI}. The duration of the IP is entirely determined by the coupling $\alpha$ and the initial value of the Hubble parameter.

\subsubsection{$1/\lambda<-36\alpha<0$}
\label{sec:late-time}

This situation is more interesting. Since $\lambda$ is negative, $w$ has the wrong sign in front of the kinetic term. The behavior of $H(t)$ is qualitatively the same as in the previous case. It decays linearly following (\ref{eq:H-IP}). Now, due to the fact that $\lambda<0$, $w$ does not enter in the oscillatory phase. Instead, it decays exponentially until the end of the IP. In this phase it can be approximated as follows
\be
w_{IP}(t)\approx w_i \exp\left[-H_i t+\frac{t^2}{72 \alpha}\right],\, t\in(t_P,t_f),\label{eq:w:IP}
\ee   
where we have assumed that $1/\vert \lambda\vert>\alpha$, $t_P$ is the Planck time, $t_f$ the time at the end of inflation defined in the caption of Fig. \ref{figpI} and $w_i$ is the initial value of the vector field. In the MFP, $w$ can be approximated by
\be
w_{MFP}(t)\approx w_{IP}(t_f)\exp\left[\sqrt{3\vert \lambda\vert}(t-t_f)\right],\,t\in(t_f,t_e),\label{eq:w:MFP}
\ee
where $t_e$ is the time at the end of the MFP. On the other hand, the following constants are exact solutions of the equations (\ref{eq:eins1})-(\ref{eq:W1})
\be
H_{dS}[t]=\sqrt{\frac{3\vert \lambda\vert}{2}},\,\,\,w_{dS}(t)=2\sqrt{\vert \lambda\vert}\label{val:asym}.
\ee

The solution (\ref{val:asym}) is a critical point of the system (\ref{eq:eins1})-(\ref{eq:W1}) and corresponds to an asymptotically stable point of the linearized system (see Appendix A). In order to estimate the duration of the MFP ($t_e$) we can use the expression (\ref{val:asym}). Since $w(t)$ grows monotonically in the MFP and (\ref{val:asym}) is an atractor, the following condition holds at $ t\approx t_e$
\be
w_{MFP}(t_e)\approx 2 \sqrt{\vert \lambda\vert}\Rightarrow t_e\approx t_f+\sqrt{\frac{\vert \lambda\vert}{3}}\log \left[\frac{2\sqrt{\vert \lambda\vert}}{w_{IP}(t_f)}\right].\label{eq:duration}
\ee

\begin{figure}
  \centering
  \includegraphics[width=0.7\textwidth]{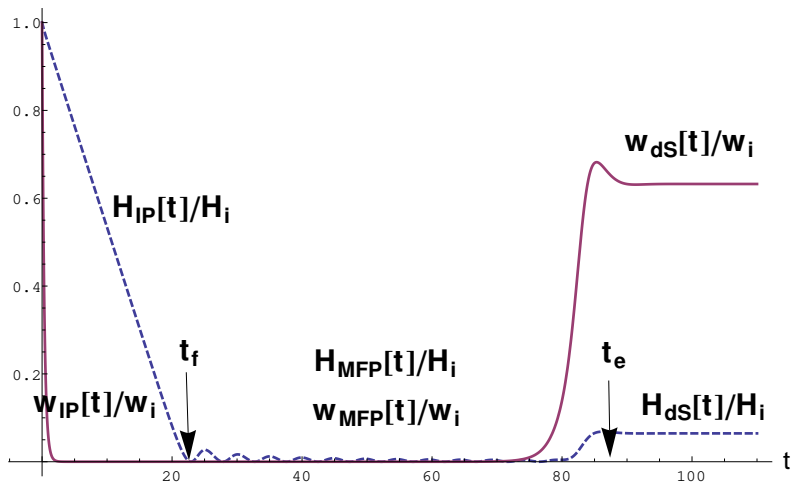}
  \caption{Qualitative behavior of the normalized Hubble parameter and $w$ vector.}
  \label{figpII}
\end{figure}

Since (\ref{val:asym}) is an exact solution of (\ref{eq:eins1})-(\ref{eq:W1}), we conclude that for $t>t_e$ the universe enters in a stable de Sitter phase. The behavior of $H(t)$ and $w(t)$ are shown qualitatively in Fig. \ref{figpII}.

\begin{figure}
  \centering
  \includegraphics[width=0.75\textwidth]{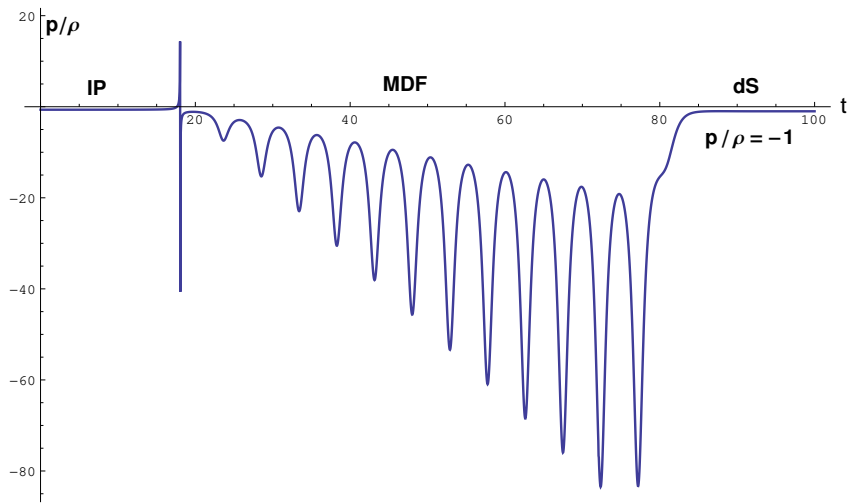}
  \caption{Equation of state. IP: inflationary phase, MFP: matter-dominated Friedmann phase and dS: de Sitter phase.}
  \label{figeos}
\end{figure}

The evolution of the equation of state $p/\rho=\omega$ is shown in Fig. \ref{figeos}. In the IP $\omega$ is approximately constant with $-1<\omega<-1/3$. After this phase it begins to oscillate to large negative values $\omega<-1$, due to the minus sign in front of the kinetic term. Finally, in the de-Sitter phase it converges to $\omega=-1$.   

\begin{figure}
  \centering
  \includegraphics[width=0.75\textwidth]{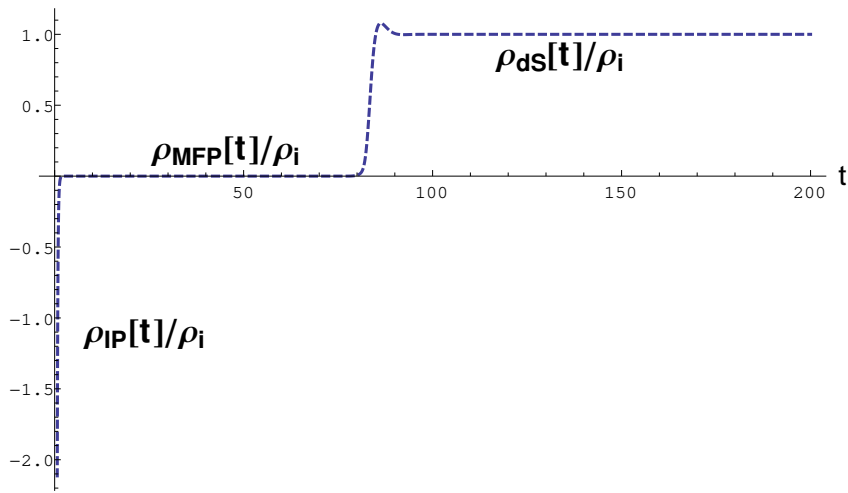}
  \caption{Normalized behavior of the defect energy density.}
  \label{figpIII}
\end{figure}

The maximum duration of the Friedmann phase depends roughly on the parameter $\lambda$ as one can see from (\ref{eq:duration}). A realistic value for $t_e$ requires the following constraints. First, in order to preserve the IP we assume first that the number of e-folds is approximately $\mathcal{N}\approx 75$, see \cite{Mijic}. Second, if the initial value $w_i$ of the Weyl vector is of the order of the coupling $\lambda$, $\vert \omega \vert^2 \approx \lambda $, the time $t_e$ for the beginning of the late-time acceleration can be computed from (\ref{eq:duration}) and gives a value $t_e\approx 25 \sqrt{3/\vert\lambda\vert}$. Therefore, if $\vert\lambda\vert$ is of the order of the square of the cosmological constant $\Lambda_c$, $t_e$ is roughly one order of magnitude bigger that its real value. Under these conditions, the model presented above describes inflation, matter-dominated Friedmann phase and late-time acceleration. It should be noted that, the potential responsible for the late-time acceleration is naturally obtained in the non-metric approach from the gravity part and no extra functions have to be chosen. Therefore, the distribution of defects, here represented by the vector field $w$, drives the accelerated stage and makes unstable the Friedmann phase. Note also that, during the late-time acceleration phase the equation of state of the defect field coincides exactly with that of a cosmological constant, $p/\rho=-1$. The qualitative behavior of the energy density of the defects in shown in Fig. \ref{figpIII}. Due to the wrong sign of the kinetic term it starts in a negative value and remains negative approximately until the end of the IP. In the MFP it takes positive values and finally stabilizes at a constant value, $\rho_d=\frac{9\vert \lambda\vert }{2}$ in the last phase. 

In the limiting case $1/\lambda=-36\alpha<0$, eqs. (\ref{EMT:limit1})-(\ref{eq:w:limit}), the universe does not enter in the MFP and the Hubble parameter goes directly from a linearly decreasing phase to a constant phase (\ref{val:asym}), leading to an continuous exponential expansion of the scale factor. The situation $\alpha<0$ is also rather uninteresting, since the Hubble parameter grows linearly leading to an exponential expansion without MFP.



\section{Conclusions}
\label{sec:conc}

We have proposed an effective description of the spacetime foam in terms of the non-metricity tensor. If such a spacetime foam is an intrinsic property of spacetime, an otherwise quite natural assumption, one should expect that its existence cannot depend on the presence or absence of matter (or radiation). Our formulation in terms of the Weyl vector suggests naturally a Maxwell term to provide the dynamics. From this point of view, the spacetime has two dynamical structures: the metric, whose dynamics is governed by the Einstein equations (or their generalization to $f(R)$ gravity), and the Weyl vector, whose dynamics is governed by Maxwell type equations (with mass term) in a curved spacetime. It is worth noting that the effect of the non-metricity is nontrivial even in flat spacetime. The trajectory of a test particle is still governed by the geodesic equation, but the connection contains now a non-vanishing part arising from the non-metricity.

It is important emphasize that the mass term, or in general the potential $V(W^\mu W_\mu)$, is naturally generated by the gravity sector $f(R)$. It is always possible to introduce by hand any potential in the Weyl vector with terms of the form $V(\widetilde \nabla_\mu g_{\mu\rho}\widetilde \nabla^\mu g^{\mu\rho})$, but this possibility increases the degree of arbitrariness of the model.

From the point of view of cosmological applications, we have pointed out that the non-metric Einstein-Hilbert action with the Maxwell term does not allow for a description of an inflationary phase. Instead, it describes a  non-accelerating and non-decelerating expansion, where the scalar factor grows linearly, as long as the potential energy dominates the kinetic one. In the opposite situation, the universe enters a radiation dominated phase. 

The situation becomes more interesting if we consider $\widetilde{R}+\alpha \widetilde{R}^2$. In this case, the behavior of the solutions depends strongly on the parameter $\alpha$, of dimension inverse mass squared, and on the coupling $\lambda$ (multiplying the Maxwell term). If $\alpha,\lambda>0$, the solutions are qualitatively similar to those of a Starobinsky-like model \cite{Starobinsky}. There is an inflationary period followed by a  matter dominated phase. If $\alpha<0$, the universe enters a de-Sitter phase that lasts forever. For $1/\lambda<-36 \alpha<0$ there are three different phases: for $t\in (t_P,t_f)$ the universe experiences a de-Sitter expansion, for $t\in (t_f,t_e)$ the universe expands as in a matter dominated phase. Finally, for $t > t_e$ the expansion begins to accelerate again. The addition of matter should not modify qualitatively our results since for large times the energy density of the defect field takes over the energy density of cold matter and radiation.
  
Regarding possible variations of the model and comparison with other available models several comments are in order:
 \begin{enumerate}
 \item As we have pointed out,  for $\lambda = 1$ and $\alpha = -1 / 36$, our model is equivalent to a scalar field $\phi=\sqrt{w^2}$ minimally coupled to $R+\alpha R^2$ which lacks the Friedmann phase. This is due to the presence of $R^2$. However, if we get rid of this term, that is, we only include a term $R\, W^\mu W_\mu$ (vector field non-minimally coupled to gravity), the Friedmann phase reappears \cite{Mukhanov-Golovnev}. But the model is still not compatible with late-time acceleration.
 \item  With an appropriated choice of initial conditions, a minimally coupled  ghost-like scalar field (with the wrong sign in front of the kinetic term) can describe inflation and Friedmann phase (driven by the higher-curvature term) and late-time acceleration (driven by the ghost field). The potential $V(\phi)$ has to be chosen such that $\pa_\phi V(\phi=\phi_0)=0$, where $\phi_0$ is the value of the $\phi$ at which the late-time acceleration begins. At this value, both $\phi_0$ and $H_0\neq 0$ are solutions of the field equations. The disadvantage of these models is that one has to choose a particular potential, which in our model is ``naturally" obtained from the gravity side.
 \item  One can also consider more general non-metric $f(R)$ gravity. This still univocally determines the potential for the Weyl vector, but also generates non-minimal couplings of the form $R^n \left(W^\mu W^\nu\right)^m$. The analysis of this kind of models is left for future investigations.
 \item  Several models previously discussed in the literature also explain in a unified way the three phases discussed here (as prominent examples see \cite{Odintsov1} for phantom inflation and \cite{Odintsov2} for Maxwell coupled to $f(R)$ and references therein). But they also require the choice of several functions (the ``metric" of the scalar field in the first case, and the non-minimal coupling function to the Maxwell term and the $f(R)$ term in the gravity side in the latter). 
\end{enumerate}
 
 Finally, the constraints obtained for $\lambda$ both in sign and magnitude, arise from the fact that, under these conditions the defect field, described by the Weyl vector, is responsible for the late-time acceleration. It should be noted that these non-canonical kinetic terms occur also in higher-derivative gravities \cite{Pollock} and  supergravities \cite{Nilles}. On the other hand, scalar fields with wrong sign in front of the kinetic term have already been considered extensively in the literature in the cosmological context \cite{Odintsov1, Caldwell, Carroll}. It may very well be that generalizations of the kinetic term for the defects, in the spirit of \cite{Odintsov2}, may change this fact. These issues are left for future work.

\vspace*{-0mm}
\section*{\hspace*{-5mm}Acknowledgments}
\vspace*{-0mm}\noindent
It is a pleasure to thank F.R. Klinkhamer and V. Emelyanov 
for useful comments on the manuscript and Z. Wang for collaborating in early stages of this project and useful discussions.

\appendix

\section{Linearized system}

The linearized system (\ref{eq:eins1})-(\ref{eq:W1}) at the point $\left(H^\star,w^\star\right)=\left(\sqrt{\frac{3\vert \lambda\vert}{2}},\,2\sqrt{\vert \lambda\vert}\right)$ has the form
\bea
\left(\begin{array}{cc}
\delta \dot{H}_1 \\ \delta \dot{w}_1\\ \delta \dot{H}\\ \delta \dot{w}
\end{array}\right)=
\left(
\begin{array}{cccc}
 -3 \sqrt{\frac{3}{2}} \sqrt{\vert \lambda\vert} & \sqrt{\vert \lambda\vert} \frac{36 \alpha +1/\vert \lambda\vert }{12 \alpha } & -6\vert \lambda\vert -\frac{1}{3 \alpha } & \frac{1+36 \alpha \vert\lambda\vert }{4
   \sqrt{6}  \alpha } \\
-\vert \lambda\vert^{3/2} 2 (36 \alpha +1/\vert \lambda\vert)& -3 \sqrt{\frac{3}{2}} \sqrt{\vert \lambda\vert} & 4 \sqrt{6}  \vert \lambda\vert^2(36 \alpha +1/ \vert \lambda\vert) & -216 \alpha \vert \lambda\vert^2   \\
 1 & 0 & 0 & 0 \\
 0 & 1 & 0 & 0
\end{array}
\right)
\left(\begin{matrix}
\delta H_1\\
 \delta w_1\\
 \delta H\\
 \delta w\\
\end{matrix}\right),\nonumber
\eea
where $H_1\equiv \dot{H}$ and $w_1\equiv \dot{w}$. The eigenvalues of the linear system can be computed in a closed form
 \bea
&& \left(\lambda_1, \, \lambda_2, \, \lambda_3, \, \lambda_4\right)=\nonumber\\
 &&\sqrt{\vert \lambda\vert}\left\{-\frac{3 \sqrt{\frac{3}{2}}}{2 }+\frac{i \sqrt{\frac{21}{2}}}{2 },-\frac{3 \sqrt{\frac{3}{2}}}{2}-\frac{i
   \sqrt{\frac{21}{2}}}{2 },-\frac{\sqrt{81 \alpha ^2+4 \alpha/\vert \lambda\vert }}{2 \sqrt{6}  \alpha }-\frac{3 \sqrt{\frac{3}{2}}}{2
   },\frac{\sqrt{81 \alpha ^2+4  \alpha/\vert \lambda\vert }}{2 \sqrt{6}  \alpha }-\frac{3 \sqrt{\frac{3}{2}}}{2 }\right\}\nonumber.
 \eea

Since we have,  $Re[\lambda_i]<0,\, i=1,2,3,4$  the critical point $\left(H^\star,w^\star\right)$ is stable. Similarly, it can be shown that the point (0,0) (corresponding to an asymptotic Friedmann universe) is unstable.

\end{document}